\documentclass[12pt]{article}

\usepackage{amssymb}
\usepackage{amsmath}
\usepackage{amscd}
\usepackage{latexsym}
\usepackage{graphicx}

\usepackage{enumerate}

\usepackage{cite}

\newcommand{\be}{\begin{equation}}
\newcommand{\ee}{\end{equation}}

\newcommand{\dlt}{\delta}

\newcommand{\bk}{{\bf k}}

\newcommand{\ba}{{\bf a}}

\newcommand{\bt}{\beta}
\newcommand{\vp}{\varphi}

\newcommand{\al}{\alpha}
\newcommand{\ra}{\rightarrow}

\newcommand{\dgr}{\dagger}
\newcommand{\lbd}{\lambda}

\newcommand{\cH}{{\cal H}}

\newcommand{\rgl}{\rangle}
\newcommand{\lgl}{\langle}

\begin{document}

\begin{center}

{\Large {\bf Measuring correlations in quantum and statistical systems} \\  [5mm]

V.I. Yukalov$^{1,2}$ and E.P. Yukalova$^3$ } \\ [3mm]

{\it $^1$Bogolubov Laboratory of Theoretical Physics, \\
Joint Institute for Nuclear Research, Dubna 141980, Russia \\ 
yukalov@theor.jinr.ru  \\[2mm]

$^2$Instituto de Fisica de S\~ao Carlos, Universidade de S\~ao Paulo, \\
CP 369, S\~ao Carlos 13560-970, S\~ao Paulo, Brazil \\ [2mm]

$^3$Laboratory of Information Technologies, \\
Joint Institute for Nuclear Research, Dubna 141980, Russia } \\ 
yukalova@theor.jinr.ru  \\ [3mm]

{\bf Corresponding author}: V.I. Yukalov \\ [2mm]

\end{center}

\vskip 2cm

\begin{abstract}

The survey discusses the problem of measuring correlations in composite quantum and 
statistical systems. After briefly recalling the standard methods of describing 
correlations, the emphasis is on the method of correlation indices. The latter method 
for quantifying the strength of correlations in composite physical systems is sufficiently 
general allowing for measuring in a unique way the correlation strength in both quantum 
and statistical systems, for static as well as for dynamic processes. The correlation 
indices, can be defined for correlation operators, reduced density matrices, and other 
operators containing information on correlations in the studied composite systems. The 
correlation indices measure all types of correlations, quantum correlations, such as 
entanglement, as well as classical correlations. They are applicable for characterizing 
static as well as dynamical processes. Examples are given of correlation indices for 
several quantum states and for spin and pseudospin systems. The quantification of 
correlations in a nonequilibrium system is exemplified by calculating the correlation 
index for a trapped Bose-Einstein condensate subject to the action of an alternating field. 
\end{abstract}

\vskip 1cm
{\parindent=0pt
{\bf Keywords}: Quantum correlations; classical correlations; entanglement; spin correlations;
dynamic correlations; Bose-Einstein condensate}

\newpage

\section{Introduction}

The notion of correlation is one of the most important notions in statistical data analysis 
and statistical physics. Different measures of correlations have been suggested, depending 
on the considered systems and setups. 
  
As a measure of correlations in analyzing numerical statistical data, one usually resorts to 
the classical Pearson coefficient \cite{Croxton_1,Dietrich_2,Aitken_3}. Suppose one observes
the sets of two random statistical variables $\xi_1$ and $\xi_2$, with the expected values 
$\overline\xi_1$ and $\overline\xi_2$ denoted as
$$
\overline\xi_i \; \equiv \;  E(\xi_i) \qquad ( i = 1,2) \; .
$$
The correlations between the variables are quantified by the Pearson coefficient
\be
\label{1}
c_{ij} \; \equiv \; 
\frac{ {\rm cov}(\xi_i,\xi_j) }{ \sqrt{{\rm var}(\xi_i) \; {\rm var}(\xi_j)} } \; ,
\ee
in which the variance and covariance, respectively, are
$$
{\rm var}(\xi_i) \; \equiv \; 
E \left[\; (\xi_i - \overline\xi_i )^2 \; \right] \; = 
\;E(\xi_i^2) - \overline\xi_i^2 \; ,
$$
$$
{\rm cov}(\xi_i,\xi_j) \; \equiv \; 
E \left[\; (\xi_i - \overline\xi_i )\; (\xi_j - \overline\xi_j ) \; \right] \; = 
\; E(\xi_i\xi_j) - \overline\xi_i \; \overline\xi_j \; ,  
$$
where $i \neq j$. By the Cauchy-Schwarz inequality
$$
 |\; {\rm cov}(\xi_i,\xi_j) \;|^2 \; \leq \; 
 {\rm var}(\xi_i) \; {\rm var}(\xi_j) \; ,
$$
it follows that $-1 \leq c_{ij} \leq 1$. 

When considering statistical systems, defined in a Hilbert space $\mathcal{H}$, and 
characterized by a statistical operator $\hat{\rho}$, the pair $\{\mathcal{H}, \hat{\rho}\}$
composes a quantum statistical ensemble. Observable quantities are represented by 
self-conjugate operators $\hat{A}$ forming a von Neuman algebra $\{\mathcal{A}\}$ of local 
observables \cite{Emch_4,Bratteli_5}. The triple $\{\mathcal{H}, \hat{\rho}, \mathcal{A}\}$ 
defines a quantum statistical system. The observable quantities are given by the averages
\be
\label{2}
 \lgl \; \hat A \; \rgl \; = \; {\rm Tr}\;\hat\rho \hat A \;  .
\ee
Correlations between two operators of observables, acting on the same Hilbert space, can 
be quantified by the quantum Pearson coefficient
\be
\label{3}
c_{AB} \; \equiv \; 
\frac{{\rm cov}(\hat A,\hat B)}{\sqrt{{\rm var}(\hat A)\;{\rm var}(\hat B)} } \;   ,
\ee
in which the variance and covariance are
$$
{\rm var}(\hat A)\; \equiv \; \lgl \; \hat A^2 \; \rgl -
\lgl \; \hat A\; \rgl^2 \; ,
$$
$$
{\rm cov}(\hat A,\hat B) \; \equiv \; \frac{1}{2} \;  \lgl \; \hat A \hat B
+ \hat B \hat A \; \rgl - \lgl \; \hat A\; \rgl \; \lgl \; \hat B\; \rgl .
$$
Note that the statistical operator $\hat{\rho}(t)$ can be a function of time $t$.

In quantum information theory, to describe quantum entanglement, that is to measure 
quantum correlations of bipartite systems in the Hilbert space 
$\mathcal{H} = \mathcal{H}_A \bigotimes \mathcal{H}_B$, with a statistical operator 
$\hat{\rho}_{AB}$, one uses several entropy-based measures 
\cite{Williams_6,Nielsen_7,Vedral_8,Keyl_9,Wilde_10,Hu_11}, such as the von Neumann 
entanglement entropy 
\be
\label{4}
S(\hat\rho_A) \; = \; - {\rm Tr}_{\cH_A}\; \hat\rho_A \; \ln \hat\rho_A \; = \;
- {\rm Tr}_{\cH_B}\; \hat\rho_B  \; \ln \hat\rho_B \;  ,
\ee
where the reduced statistical operators are
$$
\hat\rho_A \; \equiv \; {\rm Tr}_{\cH_B} \hat\rho_{AB} \; , \qquad
\hat\rho_B \; \equiv \; {\rm Tr}_{\cH_A} \hat\rho_{AB} \;   .
$$

One also uses the quantum mutual information
\be
\label{5}
I(\hat\rho_{AB}) \; \equiv \; 
S(\hat\rho_A) + S(\hat\rho_B) - S(\hat\rho_{AB}) \; ,  
\ee
where
$$
S(\hat\rho_{AB}) \; = \; - {\rm Tr}_{\cH}\; \hat\rho_{AB} \; \ln\hat\rho_{AB} \; ,
$$
and other entropy-based characteristics, such as quantum discord \cite{Hu_11}. Sometimes 
the Bell inequalities are considered \cite{Bell_12}, where quantum correlations are involved.

Generally, the existence of correlations between observables implies the nonseparability
of the correlation function of two observables, when 
\be
\label{6}
\lgl \; \hat A \hat B \; \rgl \; \not\equiv \; 
\lgl \; \hat A \; \rgl \; \lgl \; \hat B \; \rgl \;  .
\ee
On the contrary, when the condition (\ref{6}) is replaced by an equality, this means the 
absence of correlations or at least correlation weakening \cite{Bogolubov_13}.       
 
An important class of correlations in quantum systems is quantum entanglement. Different  
measures of entanglement have been discussed, mainly the measures for bipartite systems
\cite{Williams_6,Nielsen_7,Vedral_8,Keyl_9,Wilde_10,Hu_11}. Entanglement is a characteristic
of the state structure. Another characteristic is entanglement production advanced in Refs. 
\cite{Yukalov_14,Yukalov_15,Yukalov_MPL_2003}, which quantifies the amount of entanglement 
produced by the action of an operator. This measure of entanglement production has been 
studied mainly by the authors of the present survey, who considered it for the systems of 
atoms with collective atomic radiation \cite{Yukalov_LP_2004}, for Bose-Einstein condensates 
in optical lattices \cite{Yukalov_LP_2006,Yukalov_PRA_2006}, for evolution operators 
\cite{Yukalov_PRA_2015} and statistical operators \cite{Yukalov_LP_2019}, for statistical 
systems with phase transitions \cite{Yukalov_E_2020}, and for cognitive brain processes 
\cite{Yukalov_PAN_2010}.

Recently, it has been stressed \cite{Yukalov_16} that the measure of entanglement production
can be generalized to quantify the total amount of correlations in composite systems. The 
advantage of this measure is in its generality, since it can measure the correlations of all 
types, quantum as well as classical, static and dynamical, occurring in bipartite as well as in 
multipartite systems. The general measure of correlations is called the {\it correlation index}.
The difference between the correlation index and correlation function is that a correlation 
function is a qualitative characteristic of a system, showing the qualitative type of 
correlations, whether long-range or short-range, while a correlation index is a 
{\it quantitative} characteristic expressed in a numerical way. Since this measure of 
correlations is rather novel, it is described below in more detail.

\section{Correlation operators}

Let us denote the set of coordinate variables, such as spatial or momentum variables, plus
internal variables, if any, such as spin, by $x \in \mathbb{X}$. Assume that a $\sigma$-algebra
$\Sigma_X$ is given on $\mathbb{X}$. Then the pair $\{\mathbb{X}, \Sigma_X \}$ makes a
measurable space. Consider a correlation function $C_1(x,x')$ with the property
\be
\label{7}
C^*_1(x,x') \; = \; C_1(x',x) \;   .
\ee
This can be any correlation function connecting two variables. The origin of the function can
be either quantum or classical. The correlation functions can be treated as matrix elements 
of the correlation operator 
\be
\label{8}
 \hat C_1 \; = \; [\; C_1(x,x') \; ] \; , \qquad  \hat C_1^+ \; = \; \hat C_1 \; ,
\ee
which is Hermitian due to property (\ref{7}) 
 
The eigenfunctions of the correlation operator, satisfying the equation
$$
\int C_1(x,x') \; \vp_k(x') \; dx' \; = \; \lbd_k \vp_k(x) \;   ,
$$
are called natural orbitals. They form a complete orthonormal basis, 
\be
\label{9}
 \sum_k \vp_k^*(x) \; \vp_k(x') \; = \; \dlt(x-x') \; , 
\qquad
\lgl \; k \; | \; k \; \rgl \; = \; \int |\; \vp_k(x) \; |^2 dx \; = \; 1 \;  .
\ee
The column with respect to $x$ is denoted as
\be
\label{10}
|\; k \; \rgl \; \equiv \; [\; \vp_k(x) \; ] \;   .
\ee
The eigenvalues 
\be
\label{11}
\lbd_k \; = \; \lgl \; k \; | \; \hat C_1 \; | \; k \; \rgl \; = \;
\int \vp_k^*(x) \; C_1(x,x') \; \vp_k(x') \; dx dx'
\ee
play the role of occupation numbers for the state labeled by a mutli-index $k$.

The first-order correlation operator (\ref{8}) acts on the Hilbert space 
\be
\label{12}
\cH_1 \; = \; \overline {\mathcal L} \{ \; | \; k \; \rgl \; \}  
\ee
that is a closed linear envelope over the basis $\{|k \rangle\}$. The trace operation 
gives
\be
\label{13}
{\rm Tr}_{\cH_1} \hat C_1 \; = \; 
\sum_k \lgl \; k \; | \; \hat C_1 \; | \; k \; \rgl \; = \; 
\int C_1(x,x) \; dx \;   .
\ee
The norm can be defined in different forms. Here we prefer the Hermitian operator norm
\be
\label{14}
||\; \hat C_1\; || \; = \; 
\sup_\vp \; \frac{\lgl\; \vp\; |\; \hat C_1\; |\; \vp\; \rgl}{\lgl\; \vp\; |\; \vp\; \rgl} 
\; = \; \sup_k\lbd_k \;   .
\ee

Similarly, it is straightforward to introduce the second-order correlation operator
\be
\label{15}
\hat C_2 \; = \; [\; C_2(x_1,x_2,x_1',x_2') \; ] \; , \qquad
\hat C_2^+ \; = \; \hat C_2 \;   ,
\ee
whose matrix elements satisfy the property
\be
\label{16}
 C_2^*(x_1,x_2,x_1',x_2') \; = \; C_2(x_1',x_2',x_1,x_2) \;  .
\ee
This operator acts on the Hilbert space 
\be
\label{17} 
 \cH \; = \; \cH_1 \bigotimes \cH_2 \; = \; \overline {\mathcal L} \{ \; |\; k p \; \rgl \; \} \; ,
\ee
which is a closed linear envelope over the basis $\{|kp \rangle\}$ of the natural orbital
functions
\be
\label{18}
|\; k p \; \rgl \; = \; [\; \vp_{kp}(x_1,x_2) \; ] \;   .
\ee
Respectively, the eigenvalue of the correlation operator (\ref{15}) is
\be
\label{19}
\lbd_{kp} \; = \; \lgl k p \; | \; \hat C_2 \; | \; kp \; \rgl \; = \;
\int \vp_{kp}^*(x_1,x_2) \; C_2(x_1,x_2,x_1',x_2') \; \vp_{kp}(x_1',x_2') \;
dx_1 dx_2 dx_1' dx_2' \;   .
\ee

The trace of operator (\ref{15}) is   
\be
\label{20}
{\rm Tr}_\cH \hat C_2 \; = \; 
\sum_{kp} \lgl \; kp \; | \; \hat C_2 \; | \; kp \; \rgl \; = \; 
\sum_{kp} \lbd_{kp}
\ee
and its Hermitian operator norm is
\be
\label{21}
  || \; \hat C_2 \; || \; = \; \sup_{kp} \lbd_{kp}  \;  .
\ee

In the same way, it is possible to define the higher-order correlation operators
\be
\label{22}
\hat C_n \; = \; [\; C_n(x_1,x_2,\ldots,x_n,x_1',x_2',\ldots,x_n') \; ]    
\ee
acting on the Hilbert space
\be
\label{23}
 \cH \; = \; \bigotimes_{i=1}^n \cH_i \;  .
\ee

\section{Correlation indices}

The matrix elements of the correlation operator (\ref{22}) are correlation functions that,
in general, do not factorize into uncorrelated products of functions, as is shown in 
inequality (\ref{6}). For each operator (\ref{22}) one can define a factorized uncorrelated 
counterpart, whose matrix elements are represented as products of uncorrelated elements. This
factorized operator is
\be
\label{24}
 \hat C_n^\otimes \; \equiv \; 
\frac{\bigotimes_{i=1}^n \hat R_i}{({\rm Tr}_\cH \hat C_n)^{n-1} } \;  ,
\ee
which is formed by the product of reduced operators
\be
\label{25}
 \hat R_i \; \equiv \; {\rm Tr}_{\cH\setminus \cH_i} \hat C_n \;  ,
\ee
so that both operators are equally normalized 
\be
\label{26}
 {\rm Tr}_\cH \hat C_n^\otimes \; = \;  {\rm Tr}_\cH \hat C_n \;  .
\ee

The strength of correlations in a system can be measured by comparing the structure of the
correlation operator (\ref{22}) with the structure of its factorized counterpart composed
of uncorrelated operators. The measure of correlations is defined by the {\it correlation
index}
\be
\label{27}
\varkappa(\hat C_n^\otimes) \; \equiv \; 
\ln \; \frac{||\; \hat C_n\; ||}{||\; \hat C_n^\otimes \; ||} \;   .
\ee

Notice that the correlation indices can be defined for any operator $\hat{A}$, acting on 
a Hilbert space (\ref{23}), as      
\be
\label{28}
\varkappa(\hat A) \; = \; 
\ln \; \frac{||\; \hat A \; ||}{||\; \hat A^\otimes \; ||} \;   .
\ee
In particular, this can be a statistical operator $\hat{\rho}$ giving the correlation index
\be
\label{29}
\varkappa(\hat\rho) \; = \; 
\ln \; \frac{||\; \hat\rho \; ||}{||\; \hat\rho^\otimes \; ||} \;  .
\ee
Generally, the correlation index can be a function of time, when, e.g., time enters through
the statistical operator $\hat{\rho}(t)$ or through the Heisenberg operators $\hat{A}(t)$. 

Let us stress that the correlation index takes into account all types of correlations existing 
in the system, quantum as well as classical. The correlation functions used in the construction 
of the correlation index can be of any nature, either of quantum or of classical origin. Below, 
we show that the correlation indices measure such quantum correlations as entanglement. At the 
same time, they also quantify all classical correlations. For instance, let us consider a 
separable statistical operator 
\be
\label{30}
\hat\rho_{sep} \; = \; \sum_{\mu=1}^M n_\mu \bigotimes_{i=1}^N \hat\rho_\mu^{(i)} \; ,
\ee      
describing a mixed multimode system of $M$ modes, composed of $N$ parts, and acting on a Hilbert
space
\be
\label{31}
\cH \; = \; \bigotimes_{i=1}^N \cH_i \; , \qquad 
\cH_i \; = \; \overline {\mathcal L}\{\; |\; \mu \; \rgl \; \} \;   .
\ee
Here $n_\mu$ is a fractional mode population, and the normalization conditions are valid:
$$
\sum_{\mu=1}^M n_\mu \; = \; 1 \; , \qquad 
{\rm Tr}_{\cH_i} \hat\rho_\mu^{(i)} \; = \; 1 \;   .
$$
The product of the partial statistical operators 
$$
 \hat\rho_\mu^{(i)} \; = \; |\; \mu \; \rgl \lgl \; \mu \; |
$$
defines a multimode state
$$
 \bigotimes_{i=1}^N \hat\rho_\mu^{(i)} \; = \;   |\; \mu\mu \ldots\mu \; \rgl 
\lgl \; \mu \mu\ldots\mu \; | \; .
$$

The statistical operator (\ref{30}) is called separable, since it is not entangled, having the 
structure displaying no quantum correlations, such as quantum entanglement 
\cite{Nielsen_7,Vedral_8,Keyl_9,Wilde_10}, although the action of the separable operator can 
induce entangled states, when acting on disentangled factor states \cite{Yukalov_14,Yukalov_15}. 
In that sense, a separable operator can produce entangled states acting on separable wave 
functions \cite{Yukalov_16}. 

One says that the operator (\ref{30}) characterizes classical correlations. These classical 
correlations can be quantified by the correlation index $\varkappa (\hat{\rho}_{sep})$. The 
reduced single-partite operator (\ref{25}) takes the form
\be
\label{32}
 \hat\rho_i \; \equiv \; {\rm Tr}_{\cH\setminus\cH_i} \hat\rho_{sep} \; = \;
\sum_{\mu=1}^M n_\mu \; \hat\rho_\mu^{(i)} \;  ,
\ee
so that for the uncorrelated counterpart (\ref{24}), we have
\be
\label{33}
\hat\rho^\otimes \; = \; \bigotimes_{i=1}^N \hat\rho_i \; = \;
\bigotimes_{i=1}^N \left( \sum_{\mu=1}^M n_\mu \; \hat\rho_\mu^{(i)} \right) \;   .
\ee
With the norms
\be
\label{34}
||\; \hat\rho_{sep} \; || \; = \; \sup_\mu n_\mu \; , \qquad
||\; \hat\rho^\otimes\; || \; = \; \sup_\mu n_\mu^N \;   ,
\ee
we have the correlation index
\be
\label{35}
\varkappa(\hat\rho_{sep}) \; = \; 
\ln \; \frac{||\; \hat\rho_{sep}\; ||}{||\hat\rho^\otimes||} \; = \; 
( 1 - N) \sup_\mu n_\mu  
\ee
which, generally, is not zero.

\section{Density matrices}

All major properties of statistical systems, including their correlations, are incorporated 
in their reduced density matrices \cite{Coleman_16}. It is therefore reasonable to study 
correlations by considering the correlation indices of reduced density matrices.

The first-order density matrix is 
\be
\label{36}
 \rho(x,x') \; = \; {\rm Tr} \psi(x) \; \hat\rho \; \psi^\dgr(x') \; = \;
\lgl \; \psi^\dgr(x') \; \psi(x) \; \rgl \;  ,
\ee
where $\psi(x)$ are field operators and the trace is over a basis in the Fock space. It is 
admissible to treat the above expression as a matrix element of the density operator
\be
\label{37}
 \hat\rho_1 \; \equiv \; [\; \rho(x,x') \; ] \;  .
\ee
This operator is defined on the Hilbert space 
\be
\label{38}
\cH_1 \; = \; \overline {\mathcal L}\{ \; |\; k \; \rgl \; \} \; , \qquad
 |\; k \; \rgl \; = \; [\; \vp_k(x) \; ] 
\ee
that is a closed linear envelope over the basis formed of natural orbitals. The trace of 
operator (\ref{37}) reads as
\be
\label{39}
{\rm Tr}_{\cH_1} \hat\rho_1 \; = \; \sum_k \lgl \; k \; | \; \hat\rho_1 \; | \; k \; \rgl \; = \;
\int \rho(x,x) \; dx \; = \; N \;   ,   
\ee
where $N$ is the total number of particles, and its norm is
\be
\label{40}
||\; \hat\rho_1 \; || \; = \; \sup_k N_k \;   ,
\ee
with $N_k$ being the occupation number of a state $k$,
\be
\label{41}
 N_k \; = \;  \lgl \; k \; | \; \hat\rho_1 \; | \; k \; \rgl \; = \;
\int \vp_k^*(x) \; \rho(x,x') \; \vp_k(x') \; dx dx' \; .
\ee

The second-order density matrix 
\be
\label{42}
\rho_2(x_1,x_2,x_1',x_2') \; = \; 
{\rm Tr}\; \psi(x_1) \; \psi(x_2) \; \hat\rho \; \psi^\dgr(x_2') \; \psi^\dgr(x_1') \; = \;
 \lgl \; \psi^\dgr(x_2') \; \psi^\dgr(x_1') \;  \psi(x_1) \; \psi(x_2) \; \rgl \;  ,
\ee
where the trace is over the Fock space, defines the second-order density operator
\be
\label{43}
\hat\rho_2 \; \equiv \; [\; \rho_2(x_1,x_2,x_1',x_2') \; ] \; .
\ee
This operator acts on the Hilbert space defined similarly to equations (\ref{17}) and (\ref{18}). 
The trace of operator (\ref{43}) is
\be
\label{44}
 {\rm Tr}_\cH \hat\rho_2 \; = \; 
\sum_{kp} \lgl \; k p \; | \; \hat\rho_2 \; | \; kp \; \rgl \; = \; 
\int \rho_2(x_1,x_2,x_1,x_2) \; dx_1 dx_2 \;  ,
\ee
and its norm is
\be
\label{45}
|| \; \hat\rho_2 \; || \; = \; \sup_{kp} N_{kp} \;   ,
\ee
where 
\be
\label{46}
N_{kp} \; = \; \lgl \; kp \; | \; \hat\rho_2 \; | \; kp \; \rgl \; = \;
\int \vp_{kp}^*(x_1,x_2) \; \rho_2(x_1,x_2,x_1',x_2') \; \vp_{kp}(x_1',x_2') \;
dx_1 dx_2 dx_1' dx_2 ' \;   .
\ee

Introducing the reduced density operators
\be
\label{47}
\hat R_1 \; = \; {\rm Tr}_{\cH_2}\hat\rho_2 \; = \; [\; R_1(x_1,x_1') \; ] \; , 
\qquad
\hat R_2 \; = \; {\rm Tr}_{\cH_1}\hat\rho_2 \; = \; [\; R_2(x_2,x_2') \; ] \;  ,
\ee
with the matrix elements
\be
\label{48}
 R_1(x_1,x_1') \; = \; \int \rho_2(x_1,x_2,x_1',x_2) \; dx_2 \; , 
\qquad
 R_2(x_2,x_2') \; = \; \int \rho_2(x_1,x_2,x_1,x_2') \; dx_1 \; ,
\ee
we follow Sec. 3, defining the factorized operator
\be
\label{49}
 \hat\rho_2^\otimes \; = \; 
\frac{\hat R_1 \bigotimes \hat R_2}{{\rm Tr}_\cH\hat\rho_2} \;  .
\ee
 
The correlation index of $\hat{\rho}_2$ is
\be
\label{50}
 \varkappa(\hat\rho_2) \; \equiv \; 
\ln \; \frac{||\;\hat\rho_2\;||}{||\;\hat\rho_2^\otimes\;||} \;  .
\ee

\section{Quantum entanglement}

Since the correlation indices quantify all correlations, either classical or quantum, they
should also quantify as well such quantum correlations as entanglement. We show this by 
examples of bipartite systems for which entanglement can be measured by the von Neumann 
entanglement entropy.  

Let the statistical operator $\hat{\rho}$ of a bipartite system act on a Hilbert space
\be
\label{51}
\cH \; = \; \cH_1 \bigotimes \cH_2 \;   .
\ee
The reduced statistical operators are
\be
\label{52}
 \hat R_1\; \equiv \; {\rm Tr}_{\cH_2} \hat\rho \; , 
\qquad
 \hat R_2\; \equiv \; {\rm Tr}_{\cH_1} \hat\rho \; .
\ee
Let us consider the correlation indices for some concrete statistical operators and show that
these indices are nonzero together with the entanglement entropy, when the latter can be 
defined. The entanglement entropy is defined for bipartite systems as
\be
\label{53}
 S(\hat R_i) \; = \; - {\rm Tr}_{\cH_i} \hat R_i \; \ln \hat R_i \qquad ( i = 1,2) \;  .
\ee

\subsection{Einstein-Podolsky-Rosen states}

The statistical operator for a generalized Einstein-Podolsky-Rosen state reads as
\be
\label{54}
\hat\rho_{EPR} \; = \; 
(\; c_1 \; | \; 01 \; \rgl + c_2 \; | \; 10 \; \rgl \; )  
(\; c_2^* \; \lgl \; 10 \; | +c_1^* \; \lgl \; 01 \; | \; ) ,
\ee
where
$$
{\rm Tr}_\cH \hat\rho_{EPR} \; = \; 1 \; , \qquad 
|\; c_1 \; |^2 + |\; c_2 \; |^2 \; = \; 1 \;   .
$$
The reduced operators (\ref{52}) are
\be
\label{55}
\hat R_1 \; = \; |\; c_1 \; |^2 \; |\; 0 \; \rgl \lgl \; 0 \; | + 
|\; c_2 \; |^2 \; |\; 1 \; \rgl \lgl \; 1 \; | \; , 
\qquad
\hat R_2 \; = \; |\; c_2 \; |^2 \; |\; 0 \; \rgl \lgl \; 0 \; | + 
|\; c_1 \; |^2 \; |\; 1 \; \rgl \lgl \; 1 \; | \;   .
\ee
Then the uncorrelated operator is
\be
\label{56}
\hat\rho_{EPR}^\otimes \; =\; \hat R_1 \bigotimes \hat R_2 \;  .
\ee
The related norms are
$$
||\; \hat R_i \; || \; = \; 
\sup\left\{ |\; c_1 \; |^2, \; |\; c_2 \; |^2 \right\} \; ,
\qquad
||\; \hat \rho_{EPR} \; || \; = \; 
\sup\left\{ |\; c_1 \; |^2, \; |\; c_2 \; |^2 \right\} \; ,
$$
\be
\label{57}
||\; \hat \rho_{EPR}^\otimes \; || \; = \; ||\; \hat R_i \; ||^2 \; = \; 
\left( \sup\left\{ |\; c_1 \; |^2, \; |\; c_2 \; |^2 \right\} \right)^2 \; .
\ee
The correlation index becomes
\be
\label{58}
 \varkappa(\hat\rho_{EPR} ) \; = \; 
- \ln\; \sup \left\{ |\; c_1 \; |^2, \; |\; c_2 \; |^2 \right\} \;  ,
\ee
which is nonzero together with the entanglement entropy
\be
\label{59}
 S(\hat R_1) \; = \; - |\; c_1 \; |^2 \ln |\; c_1 \; |^2 -
|\; c_2 \; |^2 \ln |\; c_2 \; |^2 \;  .
\ee
The maximal values for the correlation index and entanglement entropy, when $|c_i|^2 =1/2$, 
coincide:
\be
\label{60}
\max \varkappa(\hat\rho_{EPR} ) \; = \; \max S(\hat R_i) \; = \; \ln 2 \; .
\ee

\subsection{Generalized Bell states}

The generalized $N$-partite Bell state reads as 
\be
\label{61}
 \hat\rho_B \; = \; (\; c_1\; | \; 00\ldots 0\; \rgl + c_2\; | \; 11\ldots 1\; \rgl \; )
(\; c_2^*\; \lgl \; 11\ldots 1\; | + c_1^*\; \lgl \; 00\ldots 0\; | \; ) \; .
\ee
Here
$$
{\rm Tr}_\cH \hat\rho_B \; = \; 1 \; , \qquad |\; c_1 \; |^2 + |\; c_2 \; |^2 \; = \; 1 \; .
$$
The reduced operators are 
\be
\label{62}
 \hat R_i \; = \; (\; c_1\; | \;  0\; \rgl + c_2\; | \;  1\; \rgl \; )
(\; c_2^*\; \lgl \; 1\; | + c_1^*\; \lgl \; 0\; | \; ) \;  ,
\ee
giving the uncorrelated operator
\be
\label{63}
\hat\rho_B^\otimes \; = \; \bigotimes_{i=1}^N \hat R_i \;   .
\ee
For the related norms, we have
$$
||\; \hat R_i \; || \; = \; 
\sup\left\{ |\; c_1 \; |^2, \; |\; c_2 \; |^2 \right\} \; ,
\qquad
||\; \hat \rho_B \; || \; = \; 
\sup\left\{ |\; c_1 \; |^2, \; |\; c_2 \; |^2 \right\} \; ,
$$
\be
\label{64}
||\; \hat \rho_B^\otimes \; || \; = \; ||\; \hat R_i \; ||^N \; = \; 
\left( \sup\left\{ |\; c_1 \; |^2, \; |\; c_2 \; |^2 \right\} \right)^N \;  .
\ee
This gives the correlation index
\be
\label{65}
 \varkappa(\hat\rho_B) \; = \; 
(1 - N) \; \ln \; \sup\left\{ |\; c_1\; |^2 \; , \; |\; c_2 \; |^2 \right\} \;  .
\ee

The maximal value of the correlation index
\be 
\label{66}
\max \varkappa(\hat\rho_B) \; = \; ( N - 1) \; \ln 2 \; ,
\ee
for $N=2$, this coincides with the entanglement entropy $S(\hat{R}_i) = \ln 2$ of a bipartite 
state.

\subsection{Multimode states}

The general $N$-partite state with $M$ modes has the form
\be
\label{67}
\hat\rho_{MM} \; = \; 
\left( \sum_{n=0}^{M-1} \; c_n \; | \; nn\ldots n \; \rgl \; \right)\;
\left( \sum_{n=0}^{M-1} \; c_n^* \; \lgl \; nn\ldots n \; | \; \right) \; ,
\ee
where
$$
 {\rm Tr}_\cH \hat\rho_{MM} \; = \; 1 \; , \qquad 
\sum_{n=0}^{M-1} |\; c_n \; |^2 \; = \; 1 \;  .
$$
The correlation index reads as
\be
\label{68}
\varkappa(\hat\rho_{MM} ) \; = \; ( 1 - N) \; \ln \; \sup_n |\; c_n \; |^2 \; .
\ee
The maximal correlation occurs for $|c_n| = 1/M$, when
\be
\label{69}
\max \varkappa(\hat\rho_{MM} ) \; = \; ( N - 1) \ln M \; .
\ee
The entanglement entropy for multipartite states is not defined.

\section{Hartree-Fock states}

In the Hartree-Fock state
\be
\label{70}
\hat\rho_{HF} \;  =\; \frac{1}{N!} \left( \sum_{Sym} |\; 1 2\ldots N\; \rgl \right)
\left( \sum_{Sym} \lgl \; 1 2\ldots N\; | \right) \;  ,
\ee
the subscript "Sym" implies the symmetrization or anti-symmetrization, in accordance with 
whether Bose or Fermi statistics is considered. 

Calculating the norms
\be
\label{71}
 ||\; \hat R_i \; || \; = \; \frac{1}{N}  \; , \qquad
 ||\; \hat \rho_{HF} \; || \; = \; \frac{1}{N!}  \; , \qquad
 ||\; \hat \rho_{HF}^\otimes \; || \; = \; \frac{1}{N^N}  
\ee
yields the correlation index
\be
\label{72}
 \varkappa(\hat\rho_{HF} ) \; = \; \ln\; \frac{N^N}{N!} \;  .
\ee

For a two-particle state, this gives
\be
\label{73}
 \varkappa(\hat\rho_{HF}) \; = \; \ln 2 \qquad ( N = 2) \;  ,
\ee
while for large $N$, we have
\be
\label{74}
\varkappa(\hat\rho_{HF}) \; \simeq \; N \qquad ( N \gg 1 ) \;   .
\ee

\subsection{Reduced Hartree-Fock states}

The reduced Hartree-Fock state
\be
\label{75}
 \hat\rho_n \; = \; {\rm Tr}_{\cH_{n+1}}   {\rm Tr}_{\cH_{n+2}} \ldots
{\rm Tr}_{\cH_N} \; \hat\rho_{HF}
\ee
exhibits the correlation index
\be
\label{76}
 \varkappa(\hat\rho_n) \; = \; \ln \left[ \; \frac{(N-n)!}{N!} \; N^n \; \right] \;  ,
\ee
which, for large $N$ and finite $n$, becomes
\be
\label{77}
 \varkappa(\hat\rho_n) \; \simeq \;  \frac{n(n-1)}{2N} \qquad ( N \gg 1) \; .
\ee

\section{Spin correlations}

Correlation functions can be composed of any operators of observables, in particular, of spins.
Here we show how correlation indices quantify spin correlations. For concreteness, we consider 
the spin operators $S_i^z$ of spin $1/2$, where the index $i = 1,2,\ldots,N$ enumerates the 
lattice sites ${\bf a}_i$. The pseudospin representation is also widely employed for treating
ensembles of radiating atoms \cite{Yukalov_17}, so that correlation indices can be introduced 
for pseudospin systems as well.   

Let us start with the first-order correlation operator
\be
\label{78}
 \hat C_1 \; = \; [\; C_{ij} \;] \; , \qquad 
C_{ij} \; = \; \lgl \; S_i^z \; S_j^z \; \rgl \;  ,   
\ee
whose matrix elements are the statistical averages of spin operators. The average spin is 
denoted as
\be
\label{79}
 s \;  = \; \frac{2}{N} \sum_{i=1}^N \lgl \; S_i^z \; \rgl \;  .
\ee
The correlation operator (\ref{78}) acts on the Hilbert space 
\be
\label{80}
 \cH_1 \; = \; \overline {\mathcal L}\{ \; | \; k\; \rgl \; \} \; , \qquad
 | \; k\; \rgl \; = \; [\; \vp_k(\ba_j) \; ] \; ,
\ee
which is a closed linear envelope over the natural basis of lattice functions $|k \rangle$
that are the columns with respect to the lattice sites ${\bf a}_j$, 
\be
\label{81}
 \vp_k(\ba_j) \; = \; \frac{1}{\sqrt{N}} \; e^{i\bk\cdot\ba_j} \;  .
\ee
The functions (\ref{81}) form a natural complete orthonormal basis,
$$
\sum_{i=1}^N \vp_k^*(\ba_i) \; \vp_p(\ba_i) \; = \; \dlt_{kp} \; ,
\qquad
\sum_k \vp_k^*(\ba_i) \; \vp_k(\ba_j) \; = \; \dlt_{ij} \;   .
$$

The pairs of spin operators can be represented as
$$
 S_i^z \; S_j^z \; = \; 
\frac{1}{4} \; \dlt_{ij} + ( 1 - \dlt_{ij}) \; S_i^z \; S_j^z \;  .
$$
The averages of products of spin operators at different lattice sites, assuming long-range 
interactions, can be decoupled in the mean-field approximation
\be
\label{82}
\lgl \; S_i^\al \; S_j^\bt \; \rgl \; = \; 
\lgl \; S_i^\al \; \rgl \lgl \; S_j^\bt \; \rgl \qquad ( i \neq j) \;  .
\ee
Then the correlation function $C_{ij}$ takes the form
\be
\label{83}
 C_{ij} \; = \; 
\frac{1}{4} \; \left[ \; \dlt_{ij} + ( 1 - \dlt_{ij} ) \; s^2 \; \right] \; = \; 
\frac{1}{4} \; \left[ \;  ( 1 - s^2)\; \dlt_{ij} + s^2  \; \right] \;  .
\ee

Taking into account the sums
$$
\sum_{i=1}^N \vp_k(\ba_i) \; = \; \sqrt{N} \; \dlt_{k0} \; , \qquad
\sum_k 1 \; = \; N \;   ,
$$
we get the matrix elements
\be
\label{84}
 \lgl \; k \; |\; \hat C_1 \; |\;  p \; \rgl \; = \;  \frac{1}{4}\;
\left[ \;  ( 1 - s^2)\; \dlt_{kp} + N s^2 \dlt_{k0} \dlt_{p0} \; \right] \;  .
\ee
From here the norm of $\hat{C}_1$ follows:
\be
\label{85}
||\; \hat C_1 \; || \; = \; \sup_k \;  \lgl \; k \; |\; \hat C_1 \; |\;  k \; \rgl \; = \;  
\frac{1}{4}\; \left[ \; 1 + ( N - 1 ) \; s^2 \; \right] \; .
\ee
For the trace of $\hat{C}_1$, we have
\be
\label{86}
{\rm Tr}_{\cH_1} \hat C_1 \; = \; \frac{1}{4} \; N \;   .
\ee

The second-order spin correlation operator is defined as
\be
\label{87}
 \hat C_2 \; = \; [ \; C_{ijmn} \; ] \; , \qquad
C_{ijmn} \; \equiv \; 
\lgl \; S_i^z \; S_j^z \; S_m^z \; S_n^z \; \rgl \;  .
\ee
The operator norm 
\be
\label{88}
||\; \hat C_2 \; || \; = \; \sup_{kp} \sum_{ijmn} \vp_k^*(\ba_i) \; \vp_p^*(\ba_j) \;
C_{ijmn} \; \vp_p(\ba_m) \; \vp_k(\ba_n)   
\ee
reduces to 
\be
\label{89}
 ||\; \hat C_2 \; || \; = \; \frac{1}{N^2} \sum_{ijmn} C_{ijmn} \;  ,
\ee
which, taking into account that $N\gg 1$, results in
\be
\label{90}
||\; \hat C_2 \; || \; = \; \frac{1}{16} \; \left( 6 + 6 N s^2 + N^2 s^4 \right) \;  .
\ee

The uncorrelated operator is
\be
\label{91}
\hat C_2^\otimes \;  = \; 
\frac{\hat R_1 \bigotimes \hat R_2}{{\rm Tr}_\cH \hat C_2} \; = \; 
\hat C_1 \bigotimes \hat C_1 \;  ,
\ee
where
\be
\label{92}
\hat R_1 \; = \; {\rm Tr}_{\cH_2} \hat C_2 \; = \; \frac{N}{4} \; \hat C_1 \; , 
\qquad
\hat R_2 \; = \; {\rm Tr}_{\cH_1} \hat C_2 \; = \; \frac{N}{4} \; \hat C_1 \; ,
\ee
and 
\be
\label{93}
{\rm Tr}_\cH \hat C_2 \; = \; \frac{N^2}{16} \;   .
\ee
The operator $\hat{C}_1$, acting on $\mathcal{H}_2$, is a copy of $\hat{C}_1$ acting on 
$\mathcal{H}_1$. The norm of the uncorrelated operator reads as
\be 
\label{94}
||\; \hat C_2^\otimes \; || \; = \;  ||\; \hat C_1 \; ||^2 \; = \; 
\frac{1}{16} \; ( 1 + N s^2 )^2 \;  ,
\ee
which gives the correlation index
\be
\label{95}
 \varkappa(\hat C_2) \; = \; 
\ln \; \frac{||\; \hat C_2 \; ||}{||\; \hat C_2^\otimes \; ||} =
\ln \; \frac{6 + 6Ns^2+N^2s^4}{(1+Ns^2)^2} \;   .
\ee

The correlation index (\ref{95}) is maximal, when there is no order in the system, 
\be
\label{96}
 \varkappa(\hat C_2) \; = \; \ln 6 \qquad ( s = 0 ) \;   .
\ee
With the rising order, under a large number of sites, the correlation index diminishes, 
tending to zero under complete order,
\be
\label{97}   
  \varkappa(\hat C_2) \; \simeq \; 0 \qquad ( Ns^2 \ra \infty ) \;  .
\ee
This behavior is easily understood remembering that, under complete order in a system,
the averages of spin operators decouple, thus showing the absence of correlations in the 
sense of condition (\ref{6}), while in the paramagnetic phase, which is in a random state,
the mean-field decoupling is not allowed, hence demonstrating strong correlations.

\section{Correlation dynamics}

When correlation functions or statistical operators describe nonequilibrium systems, 
correlation indices depend on time. As an example, let us consider the case of a trapped
Bose-Einstein condensate subject to the action of an alternating field modulating the 
trapping potential \cite{Yukalov_PRA_1997,Yukalov_18,Yukalov_19}. 

Let a Bose-Einstein condensate be trapped in a potential possessing several potential wells.
If the number of wells is large and they are periodically distributed in space, we have a
kind of a lattice. The simplest case corresponds to the setup of two potential wells forming
a double-well potential. The wells are assumed to be sufficiently deep, so that atomic 
transitions between the wells are suppressed. At low temperature, almost all atoms can be 
Bose-condensed, piling down to the lowest energy levels. Suppose the trap is subject to a 
modulating field synchronously shaking the trap wells with the frequency being in 
resonance with the transition frequency between the lowest energy level and an excited level. 
Under this kind of resonance, only these two levels are involved in the dynamics. The 
statistical operator for the considered case of two wells and two modes have the Bell form
\be
\label{98}
 \hat\rho \; = \; ( \; c_1 \; | \; 00 \; \rgl + c_2 \; | \; 11 \; \rgl \; )
  ( \; c_2^* \; \lgl \; 11 \; | + c_1^* \; \lgl \; 00 \; | \; ) \; ,
\ee
with the condition
$$
|\; c_1 \; |^2 + |\; c_2 \; |^2 \; = \; 1 \; , \qquad 
{\rm Tr} \hat\rho \; = \; 1 \;  .
$$
This Bell state (\ref{98}) is a particular case of the generalized Bell state (\ref{61}) with 
the fractional mode populations $|c_1|^2$ and $|c_2|^2$. 

If the two-well trap is shaken so that in both wells the same modes are excited synchronously, 
then we realize the state (\ref{98}). Dynamics of the modes in each of the wells is described
by the nonlinear Shr\"{o}dinger equation. The condensate wave function can be expanded over
the modes, finally obtaining the temporal dependence for the mode populations 
\cite{Yukalov_PRA_1997,Yukalov_18,Yukalov_19}. 

It is convenient to introduce the fractional population difference
\be
\label{99}
 z \; = \; |\; c_1 \; |^2 - |\; c_2 \; |^2 \; .
\ee
Then the mode populations are expressed through their difference as
\be
\label{100}
|\; c_1 \; |^2  \; = \; \frac{1-z}{2} \; , \qquad 
|\; c_2 \; |^2  \; = \; \frac{1+z}{2} \;  .
\ee
The coefficients $c_i$ are represented in the form
\be
\label{101}
 c_1 \; = \; |\; c_1 \; | \; e^{i\vp_1} \; , \qquad
 c_2 \; = \; |\; c_2 \; | \; e^{i\vp_2} \; ,
\ee
with $\varphi_i$ being the real phases. Introducing the phase difference 
\be
\label{102} 
\vp \; \equiv \; \vp_2 - \vp_1 \;,   
\ee
we have
\be
\label{103}
c_1^* \; c_2 \; = \; |\; c_1 \; c_2 \; | \; e^{i\vp} \;   .
\ee

Using the above notations in the nonlinear Schr\"{o}dinger equation, describing the 
Bose-Einstein condensate, makes it possible \cite{Yukalov_PRA_1997,Yukalov_18,Yukalov_19} 
to derive the equations of motion for the population difference $z$ and phase difference 
$\varphi$, which in dimensionless units are
\be
\label{104}
\frac{dz}{dt} \; = \; - b\; \sqrt{1-z^2} \; \sin\vp \; , \qquad
\frac{d\vp}{dt} \; = \; z + \frac{bz}{\sqrt{1-z^2}} \; \cos\vp \;   .
\ee
The parameter $b$ describes the ratio of the modulating field amplitude to the strength of  
atomic interactions. This parameter $b$ can be called the pumping parameter. 

The correlation index for the statistical operator (\ref{98}) is calculated as is explained 
in Sec. 5, yielding
\be
\label{105}
 \varkappa(\hat\rho) \; = \; 
- \ln \; \sup \left\{ |\; c_1\; |^2 ; \; |\; c_2\; |^2 \right\} \; .
\ee
In view of representation (\ref{100}), this gives the expression
\be 
\label{106}
 \varkappa(\hat\rho) \; = \; 
- \ln \; \sup \left\{ \frac{1-z}{2} \; ; \; \frac{1+z}{2} \right\} \;  .
\ee
  
Equations (\ref{104}) describe the time dependence of the population difference $z = z(t)$, 
substituting which into expression (\ref{106}) defines the temporal behavior of the 
correlation index. At the initial moment of time, the system is in its ground state, with
the initial condition 
\be
\label{107}
 z(0) \; = \; -1 \; , \qquad \vp(0) \; = \; 0 \;  .
\ee
The transfer to the excited state starts with switching on the modulation field characterized
by the pumping parameter $b$. 

The population-difference dynamics exhibits two main regimes of behavior. At low pumping-field 
amplitude, when $0 < b < 1$, there is the Josephson regime, while at high pumping-field 
amplitude, when $b > 1$, the Rabi regime is realized \cite{Yukalov_PRA_1997,Yukalov_19}. 
Substituting the solution for $z(t)$ into the correlation index (\ref{106}) leads to the 
behavior of the correlation index shown in Figs. 1 and 2 for the Josephson regime and in 
Fig. 3, for the Rabi regime. By varying the pumping parameter $b$ one can regulate the 
amplitude and periodicity of the correlation index oscillations. The maxima of the correlation 
index are achieved at $z = 0$, where the correlation index reaches the value $\ln 2$. The 
minima of the correlation index occur at $z = \pm 1$, where the correlation index becomes 
zero, so that the system is the most ordered but least correlated.   

\begin{figure}[ht]
\centerline{
\includegraphics[width=8.5cm]{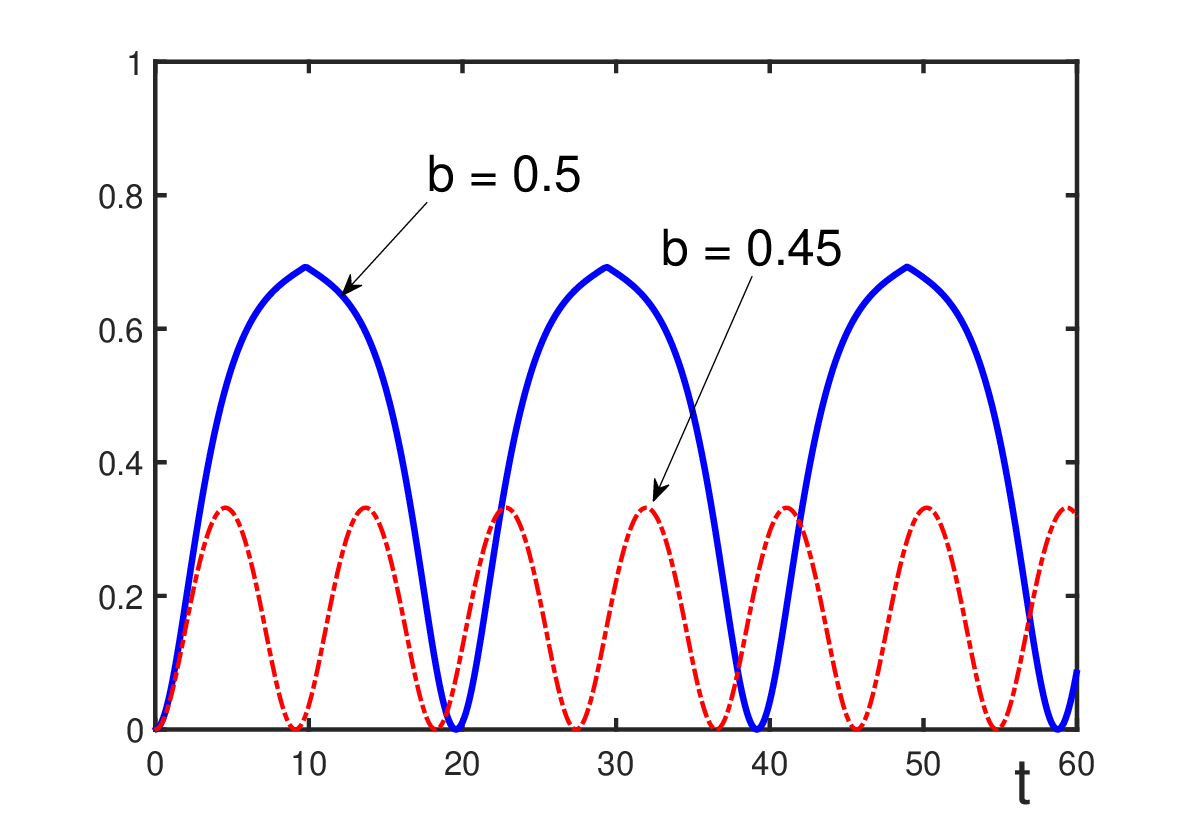} }
\caption{\small
Correlation index $\varkappa(\hat\rho)$ as a function of
dimensionless time for the Josephson regime with the pumping
parameter in the range $0<b\leq 1/2$.
}
\label{fig:Fig.1}
\end{figure}

\begin{figure}[ht]
\centerline{
\includegraphics[width=8.5cm]{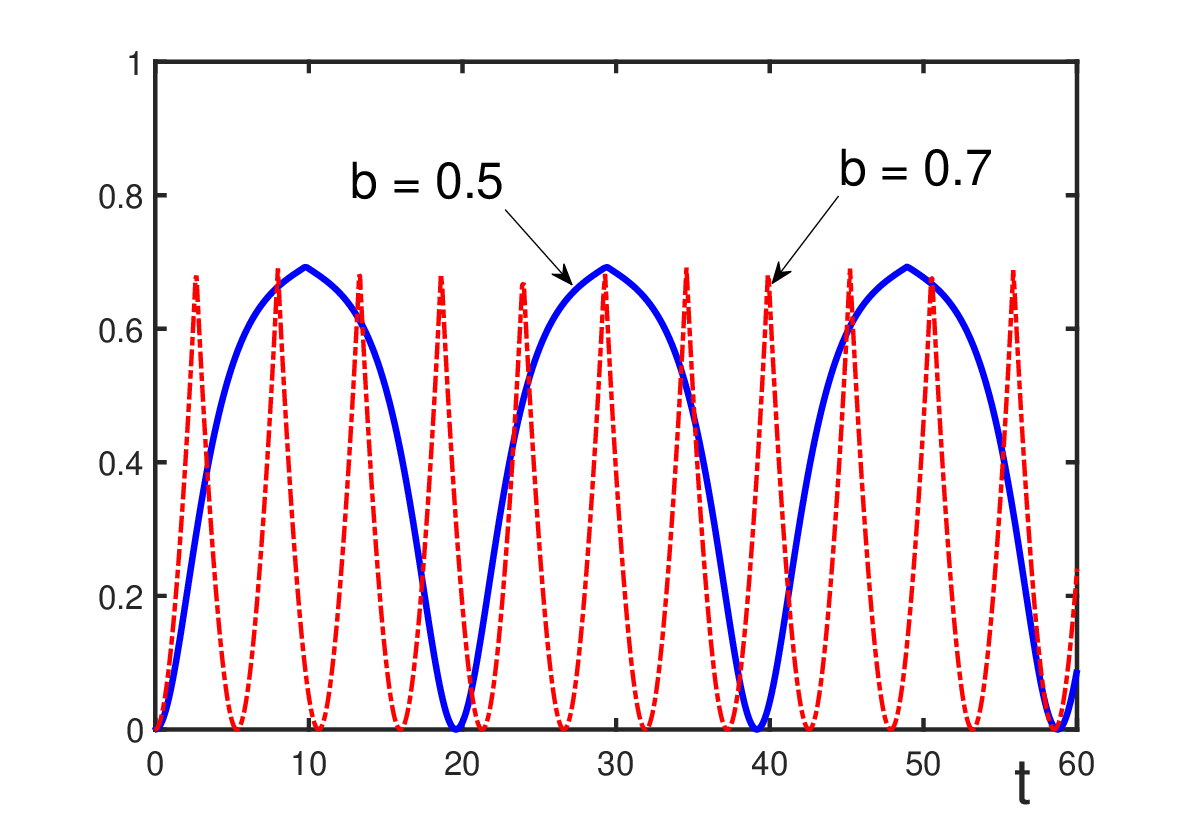} }
\caption{\small
Correlation index $\varkappa(\hat\rho)$ as a function of
dimensionless time for the Josephson regime with the pumping
parameter in the range $1/2\leq b < 1 $.
}
\label{fig:Fig.2}
\end{figure}

\begin{figure}[ht]
\centerline{
\includegraphics[width=8.5cm]{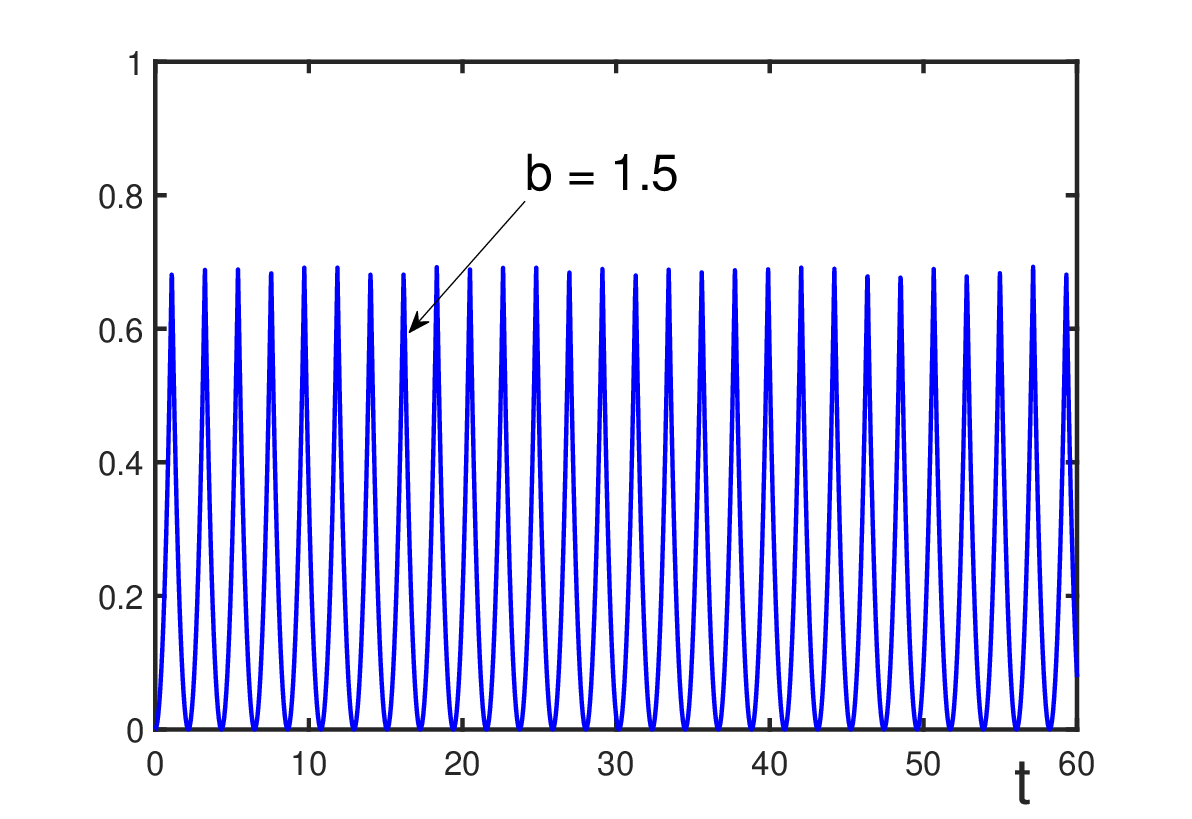} }
\caption{\small
Correlation index $\varkappa(\hat\rho)$ as a function of
dimensionless time for the Rabi regime with the pumping 
parameter in the range $b > 1 $.
}
\label{fig:Fig.3}
\end{figure}

\section{Conclusion}

The methods of measuring correlations in composite systems are briefly reviewed. The emphasis
is made on the recently developed method of correlation indices. This measure provides not
merely a qualitative characteristic of correlations, being, for instance, long-range or 
short-range, but quantifies correlations numerically. 
 
Correlation indices are the characteristics of operators containing information on correlations  
in the system. Such information on correlations is endowed, e.g., in correlation operators 
and reduced density operators. The correlation indices take into consideration the total 
correlations in the system, quantum as well as classical, equilibrium and nonequilibrium. 
The considered systems can be of any nature, composed of either particles, spins, or pseudospins. 

Technically, the calculation of correlation indices requires the knowledge of correlation 
functions or reduced density matrices. In some cases, mean-field type approximations are 
acceptable. In other cases, one needs more refined approximations or numerical calculations. 
For example, in recent years, the multiconfigurational time-dependent Hartree approach 
\cite{Alon_20,Sakman_21,Lode_22,Alon_23} became widely popular. This approach is very 
effective in studying the structure and temporal behavior of reduced density matrices. 
It would be interesting to apply this approach for calculating the correlation indices for 
different systems. The advantage of describing composite systems by using correlation indices 
is in the ability of quantifying the strength of correlations, thus obtaining not just a 
qualitative, but a quantitative characteristic of correlation strength.            

Finally, measuring correlations with correlation indices can also be done in statistical 
analysis of big data sets for discovering, quantifying, and predicting relationships between 
the sequences of observed variables in different fields. This is because the sets of 
statistical data, as is mentioned in the Introduction, can be considered as functions of 
discrete variables. Therefore, quantifying the data sets with correlation indices can be 
used for data reduction, diagnostic purposes, and for identifying patterns by measuring 
the strength of associations in various applications, including the analysis and feature 
selection in machine learning. 

\vskip 1cm
{\parindent=0pt
{\bf Authors' contribution}: Conceptualization, V.Y. and E.P.; Methodology, V.Y. and E.P.; 
Software, E.P.; Validation, V.Y. and E.P.; Formal Analysis, V.Y.; Investigation, V.Y. and E.P.; 
Writing – Original Draft Preparation, V.Y.; Writing – Review $\&$ Editing, V.Y. and E.P.; 
Visualization, E.P.; Supervision, V.Y. 

\vskip 2mm

{\bf Funding}: This research received no external funding

\vskip 2mm

{\bf Conflicts of Interest}: The authors declare no conflict of interest.

\vskip 2mm

{\bf Data availability status}: No new data were created or analyzed in this study. Data 
sharing is not applicable to this article.
}


\vskip 2cm

\begin{thebibliography}{99}

\bibitem{Croxton_1}
Croxton F E, Cowden D J and Klein S 1968
{\it Applied General Statistics} 
(Pitman: London)

\bibitem{Dietrich_2}
Dietrich C F 1991 
{\it Uncertainty, Calibration, and Probability: The Statistics of Scientific and Industrial 
Measurement} 
(Hilger: Bristol)

\bibitem{Aitken_3}
Aitken A  C 1962
{\it Statistical Mathematics} 
(Oliver $\&$ Boyd: London) 

\bibitem{Emch_4}
Emch G G 1972
{\it Algebraic Methods in Statistical Mechanics and Quantum Field Theory}
(Wiley: New York)

\bibitem{Bratteli_5}
Bratteli O and Robinson D W 1979
{\it Operator Algebra and Quantum Statistical Mechanics}
(Springer: New York)

\bibitem{Williams_6} 
Williams C P and Clearwater S H 1998 
{\it Explorations in Quantum Computing} 
(Springer: New York)

\bibitem{Nielsen_7}
Nielsen M A and Chuang I L 2000 
{\it Quantum Computation and Quantum Information} 
(Cambridge University Press: Cambridge)

\bibitem{Vedral_8}
Vedral V 2002 
The role of relative entropy in quantum information theory 
{\it Rev. Mod. Phys.} {\bf 74} 197--234

\bibitem{Keyl_9}
Keyl M 2002 
Fundamentals of quantum information theory 
{\it Phys. Rep.} {\bf 369} 431--548

\bibitem{Wilde_10}
Wilde M 2013 
{\it Quantum Information Theory} 
(Cambridge University Press: Cambridge)

\bibitem{Hu_11}
Hu M L, Hu X, Wang J, Peng Y, Zhang Y R, Fan H 2018
Quantum coherence and geometric quantum discord
{\it Phys. Rep.} {\bf 762} 1--100

\bibitem{Bell_12}
Bell J S 1987 
{\it Speakable and Unspeakable in Quantum Mechanics} 
(Cambridge University Press: Cambridge) 

\bibitem{Bogolubov_13}
Bogolubov N N 2015 
{\it Quantum Statistical Mechanics} 
(World Scientific: Singapore)

\bibitem{Yukalov_14}
Yukalov V I 2003
Entanglement measure for composite systems
{\it Phys. Rev. Lett.} {\bf 90} 167905

\bibitem{Yukalov_15}
Yukalov V I 2003
Quantifying entanglement production of quantum operations
{\it Phys. Rev. A} {\bf 68} 022109

\bibitem{Yukalov_MPL_2003}
Yukalov V I 2003
Evolutional entanglement in nonequilibrium processes
{\it Mod. Phys. Lett. B} {\bf 17} 95--103 

\bibitem{Yukalov_LP_2004}
Yukalov V I 2004
Entanglement production under collective radiation
{\it Laser Phys} {\bf 14} 1403--1414 

\bibitem{Yukalov_LP_2006}
Yukalov V I and Yukalova E P 2006
Entanglement production with multimode Bose-Einstein condensates in optical lattices
{\it Laser Phys} {\bf 16} 354--359 

\bibitem{Yukalov_PRA_2006}
Yukalov V I and Yukalova E P 2006
Regulating entanglement production in multitrap Bose-Einstein condensates
{\it Phys Rev A} {\bf 73} 022335 

\bibitem{Yukalov_PRA_2015}
Yukalov V I and Yukalova E P 2015
Evolutional entanglement production
{\it Phys Rev A} {\bf 92} 052121 

\bibitem{Yukalov_LP_2019}
Yukalov V I, Yukalova E P and Yurovsky V A 2019 
Entanglement production by statistical operators
{\it Laser Phys} {\bf 29} 065502 

\bibitem{Yukalov_E_2020}
Yukalov V I 2020
Order indices and entanglement production in quantum systems
{\it Entropy} {\bf 22} 565 

\bibitem{Yukalov_PAN_2010}
Yukalov V I and Sornette D 2010
Entanglement production in quantum decision making
{\it Phys At Nucl} {\bf 73} 559--562 

\bibitem{Yukalov_16}
Yukalov V I and Yukalova E P 2025
Measure of entanglement production by quantum operations
{\it Phys Part Nucl} {\bf 56} 999--1003 

\bibitem{Coleman_16}
Coleman A J and Yukalov V I 2000
{\it Reduced Density Matrices}
(Springer: Berlin) 

\bibitem{Yukalov_17}
Yukalov V I 2014
Coherent dynamics of radiating atomic systems in pseudospin representation
{\it Laser Phys.} {\bf 24} 094015 

\bibitem{Yukalov_PRA_1997}
Yukalov V I, Yukalova E P and Bagnato V S 1997
Non-ground-state Bose-Einstein condensates of trapped atoms
{\it Phys Rev A} {\bf 56} 4845--4854 

\bibitem{Yukalov_18}
Yukalov V I, Yukalova E P and Bagnato V S 2023
Trapped Bose-Einstein condensates with nonlinear coherent modes
{\it Laser Phys} {\it 33} 123001 

\bibitem{Yukalov_19} 
Yukalov V I and Yukalova E P 2025
Dynamical transitions in trapped superfluids excited by alternating fields
{\it Physics} {\bf 7} 41

\bibitem{Alon_20}
Alon A E, Streltsov A I and Cederbaum L S 2008
Multiconfigurational time-dependent Hartree method for bosons: Many-body dynamics of bosonic 
systems
{\it Phys Rev A} {\bf 77} 033613

\bibitem{Sakman_21}
Sakmann K, Streltsov A I, Alon O E and Cederbaum L S 2008
Reduced density matrices and coherence of trapped interacting bosons
{\it Phys. Rev. A} {\bf 78} 023615

\bibitem{Lode_22}
Lode A U J, L\'{e}v\"{e}que C, Madsen L B, Streltsov A I and Alon O E 2020
Multiconfigurational time-dependent Hartree approaches for indistinguishable particles
{\it Rev. Mod. Phys.} {\bf 92} 011001 

\bibitem{Alon_23}
Alon O E and Cederbaum L S 2025
Fragmentation of a trapped multi-species bosonic mixture
{\it Physics} {\bf 7} 38 

\end{thebibliography}
\end{document}